\documentclass[twocolumn,showpacs,floatfix,aps,nofootinbib]{revtex4}
\usepackage[dvips]{graphics,color}
\usepackage{epsfig}\usepackage{float}
\usepackage{bm}
\RequirePackage{amssymb}
\RequirePackage{amsmath}
\usepackage{makeidx}
\newcommand{\beq}{\begin{eqnarray}}\newcommand{\benu}{\begin{enumerate}}\newcommand{\enu}{\end{enumerate}}
\newcommand{\eeq}{\end{eqnarray}}
\usepackage[brazil]{babel}
\usepackage[latin1]{inputenc}
\usepackage{graphicx}
\usepackage{indentfirst}
\usepackage{color}
\newcommand{\be}{\begin{equation}}
\newcommand{\ee}{\end{equation}}

\newcommand{\ba}{\begin{eqnarray}}
\newcommand{\ea}{\end{eqnarray}}

\def\0{\bm0}
\def\vp{\bm\varphi}

\def\kb{\bm\kappa}

\def\s{\bm\sigma}
\def\p{\bm{p}}
\begin{document}

\title{Hawking radiation from 
Elko particles tunnelling across black strings horizon}
\author{Rold\~ao da Rocha}
\email{roldao.rocha@ufabc.edu.br} \affiliation{
Centro de Matem\'atica, Computa\c c\~ao e Cogni\c c\~ao,
Universidade Federal do ABC, 09210-580, Santo Andr\'e - SP, Brazil.}
\affiliation{International School for Advanced Studies (SISSA), Via Bonomea 265, 34136 Trieste, Italy.}
\author{J. M. Hoff da Silva}
\email{hoff@feg.unesp.br} \affiliation{UNESP - Campus de Guaratinguet\'a - DFQ, 12516-410,  Guaratinguet\'a - SP,
Brazil.}

\begin{abstract}
We apply the tunnelling method 
for the emission and absorption of Elko
particles in the event horizon of a black string solution. We  show  that Elko particles are emitted at the expected Hawking temperature from black strings, but with a quite different signature with respect to the Dirac particles. We employ the Hamilton-Jacobi 
 technique  to black hole tunnelling, by applying the WKB approximation to the coupled system of Dirac-like equations
 governing the Elko particle dynamics. As a typical signature, different Elko particles are shown to produce the same standard Hawking temperature for black strings.  {}{However we prove that they present the same probability irrespective of  outgoing or ingoing the black hole horizon. It provides a typical signature for mass dimension one fermions, that is different from the mass dimension three halves fermions inherent to Dirac particles, as different Dirac spinor fields have distinct inward and outward 
probability of tunnelling.}\end{abstract}
\maketitle

Black hole tunnelling procedures have been placed as prominent   methods of calculating the  temperature  of black holes \cite{1,111,2,3,4,5,6,7,8,9,
kernergodel,10,11,12}.
The tunnelling method  imparts a dynamical model describing the black hole radiation, and has been applied to a plenty of black holes, both for the tunnelling of Dirac particles \cite{10,11,12,k0,k1,ah2,ah4} and scalar  particles as well \cite{1,111}.
The first black hole tunnelling method \cite{2} succeeded from the seminal result in  \cite{1,111}. An alternative  technique  to study the  tunnelling in black hole is the Hamilton-Jacobi one \cite{3} by settling on a suitable ansatz for the action. This method was further extended, by applying the WKB approximation to the Dirac equation \cite{12,k0,k1,ah2,ah4}.
The black hole tunnelling method has strong points with respect to other routines of calculating temperature, and  can be successfully further applied to other black holes    \cite{7,8,k0, 10, vaidya, kernergodel}.  The tunnelling method provides a natural framework to study the black hole radiation, where a particle trails a path from the inside of the black hole to the outside, which is  a banned possibility from the classical point of view. By energy conservation, the black hole radius constricts as a function of the energy of the outgoing particle, hence the particle  provides its own tunnelling barrier \cite{k0, k1}. 

 {}{A quantum WKB approach was used to  compute the corrections to the Hawking temperature and Bekenstein-Hawking entropy for the Schwarzschild black hole, modifying  the Schwarzschild metric which takes into account effects of quantum corrections \cite{Banerjee:2008fz,Banerjee:2008cf,Banerjee:2009wb,Banerjee:2008ry}. Furthermore, the black hole area was shown to have a  lower bound \cite{Banerjee:2010be} in tunnelling formalism. The chirality condition was likewise introduced to  unify the anomaly and the tunnelling formalisms for deriving the Hawking effect \cite{Banerjee:2008sn}, and the Hawking radiation from the black hole, both in Ho$\check{\rm r}$ava-Lifshitz and Einstein-Gauss-Bonnet gravities, was discussed in \cite{Majhi:2009xh,Banerjee:2009pf}. Important achievements have been also accomplished in, e. g., \cite{Banerjee:2008gc} in a non-commutative framework.}

{}{The tunnelling method has been employed to provide Hawking radiation due to photon and gravitino tunnelling \cite{Majhi:2009uk}. }Moreover, this method was extended to model the emission of spin-1/2 fermions, {}{and the Hawking  radiation  was deeply analyzed in \cite{Majhi:2008gi} as tunnelling of Dirac particle throughout an event horizon, where quantum corrections in the single particle action are proportional to the usual semiclassical contribution and the modifications to the Hawking temperature and Bekenstein-Hawking entropy were derived for the Schwarzschild black hole.} When fermions of  spin-1/2 are taken into account, due to the fact that there are particles with the spin in any direction, the effect of the spin of each type of fermion countervails, and thus the lowest WKB order provides that the rotation of the black hole is unmodified. 
{}{The authors in \cite{6} argued that  the probability of emission of a particle approaches zero when its energy becomes of the order of the mass of the emitting black hole.  Consequently, the event horizon decreases \cite{6,kernergodel,k0}, and the usual approximations used} in the literature \cite{1,111,2,3,4,5,6,7,8,9,
kernergodel,10,11,12,k0,k1,ah2,ah4} remain to be adopted here.

Elko (dark) spinor fields (dual-helicity eigenspinors of
the charge conjugation operator \cite{allu}) are  spin-1/2  fermions of mass dimension one, with  novel features that make them capable to  incorporate both the Very Special Relativity (VSR) paradigm \cite{horv1} and the dark matter description as well \cite{allu,allu1,al2,al3,horv1}.
Moreover, an Elko spinor mass generation mechanism has been introduced in  \cite{alex},  by a natural coupling to the {kink} solution of a $\lambda \phi^{4}$ field theory. 
It provides exotic couplings among scalar field topological solutions and Elko spinor fields   \cite{alex,daSilva:2012wp}. 
Some attempts to detect Elko at the LHC have been proposed \cite{plb,plb1,plb2}, as well as   important applications to cosmology  have been widely investigated \cite{fab,jhep,saulo,sh,bur1,bur2}. {}{Not merely in quantum field theory, and supersymmetry \cite{wund}, but additionally the Einstein-Hilbert, the Einstein-Palatini, and the Holst actions were shown to be derived from the quadratic spinor Lagrangian, when Elko spinor fields are considered \cite{ro11,ro12}.}

 The tunnelling method is used in this paper to model Elko particles emission and absorption from black strings. We show  that Elko particles are emitted at the expected Hawking temperature from  black holes and black strings, providing further evidence for the universality of black hole radiation \cite{k0,k1,kernergodel},  however with a specific signature that is different from Dirac particles.  {}{In fact, we shall prove that Elko particles behave contrastively  from Dirac particles, that present different inward and outward 
probability of tunnelling --- depending on a relationship between the spinor components \cite{ah2,ah4}. In fact, we shall show that the four distinct Elko particles, being  
eigenspinors of the charge conjugation 
operator with dual helicity, manifest the property of 
 presenting the same equations for tunnelling, and consequently  the same inward and outward probability of 
 tunnelling. Moreover,  the standard  Hawking temperature for black  strings is obtained in this context. } The results presented in this paper for Elko particles differ from Dirac particles, as naturally Elko particles are fields presenting mass dimension one \cite{allu,allu1,al2,al3,jhep}.

String theory has solutions describing extra-dimensional extended objects surrounded by event horizons, namely black strings. These solutions can have unusual causal structure, and provide some insight into the properties of singularities in string theory. 
Black strings have been studied in the context of supergravity theories, topological defects and low energy string theories \cite{horo,ander,4658} and from pure gravitational context in  \cite{Chamblin:1999by,clark}, as well as in some realistic generalizations \cite{plbb,pl1,pl2}.

Einstein equations has solutions  
\cite{lemos95a, LZ}
\begin{equation}
ds^{2}=-f(r)dt^{2}+\frac{1}{f(r)}dr^{2}+r^{2}d\theta^{2}+\alpha ^{2}r^{2}dz^{2},  \label{bsss}
\end{equation}
where  $\Lambda_{(4)}=-3\alpha^2$ denotes the cosmological
constant, $M$ is associated to the ADM mass density of the black
string, and $
f\left( r\right) = \left( \alpha ^{2}r^{2}-\frac{M}{r}\right).$ The event horizon of the black hole is provided forthwith by:
\begin{equation}
\tilde{r}=\left( \frac{M}{\alpha ^{2}}\right) ^{{1}/{3}}.
\nonumber
\end{equation}
Also, this solution was discussed in \cite{CZ} in the context of Einstein-Maxwell gravity.

In order to analyze the tunnelling of fermions throughout the black string horizon, we 
depart from the usual mass dimension 3/2 fermions, and shall investigate the role that Elko particles play in this context. To accomplish it, the basic features of Elko
 particles  are briefly
revisited \cite{allu,allu1,crevice}. Elko spinor fields $\lambda(p^\mu)$ are eigenspinors of the charge
conjugation operator $C$, namely, $C\lambda(p^\mu)=\pm \lambda(p^\mu)$ (here the momentum space is used just to fix the notation). The Weyl representation of $\gamma^{\mu}$ is used hereupon. 
The plus [minus] sign regards {self-conjugate},  [{anti self-conjugate}]  spinor fields, denoted by $\lambda^{S}(p^\mu)$ [$\lambda^{A}(p^\mu)$]. Explicitly, once the rest spinors  $\lambda(k^\mu)$ are obtained, for an arbitrary $p^\mu$ it yields 
\begin{equation}
	\lambda(p^\mu) = e^{i \kb\cdot\vp} \lambda(k^\mu),   \label{boost}
\end{equation} where $
		k^\mu = \left(m,\lim_{p\rightarrow 0}\frac{\p}{p}\right),$ for $p = \vert\p\vert.$ 
The boost operator in 
(\ref{boost}) is provided by \cite{crevice}
\begin{eqnarray}
 e^{i \kb\cdot\vp} = \sqrt{\frac{E + m }{2 m}}
						{\rm diag}
						\left(\mathbb{I} + \frac{\s\cdot\p}{E +m},  \mathbb{I} - \frac{\s\cdot\p}{E +m} 
						\right).  \nonumber
\end{eqnarray}
The $\phi_{}(k^\mu)$ are defined to be  eigenspinors of  the helicity operator 
$\s\cdot\hat \p$:
\begin{equation}
\s\cdot\hat \p\, \phi_{}^\pm(k^\mu) = \pm \phi_{}^\pm(k^\mu)\nonumber
\end{equation}
where $\hat \p = (\sin\theta\cos\phi, \sin\theta\sin\phi,\cos\theta)$, and the phases are employed \cite{allu,allu1,crevice} such that 
\begin{eqnarray}
\phi^+_{}(k^\mu) &=& \sqrt{m} \left(
									\begin{array}{c}
									\cos\left(\frac{\theta}{2}\right)e^{- i \phi/2}\\
									\sin\left(\frac{\theta}{2}\right)e^{+i \phi/2}
											\end{array}
									\right)\,, \label{phim}\\ 
\phi^-_{}(k^\mu) &=& \sqrt{m} \left(
									\begin{array}{c}
									-\sin\left(\frac{\theta}{2}\right)e^{- i \phi/2}\\
									\cos\left(\frac{\theta}{2}\right)e^{+i \phi/2}
											\end{array}
									\right)\,. 	\label{phime}					\end{eqnarray}
Elko spinor fields  $\lambda(k^\mu)$ are defined by
\begin{eqnarray}
 \lambda^S_\pm(k^\mu) & =& \left(
					\begin{array}{c}
					i \Theta\left[\phi_{}^\pm(k^\mu)\right]^\ast\\
								\phi_{}^\pm(k^\mu)
					\end{array}
					\right) 
					\label{pppm} 									\\ \lambda^A_\pm(k^\mu) &=& \pm\left(
							\begin{array}{c}
								- i \Theta\left[\phi_{}^\mp(k^\mu)\right]^\ast\\
								\phi_{}^\mp(k^\mu)
												\end{array}
												\right)
			\label{ppm}
\end{eqnarray}
where the $\Theta$ denotes the Wigner time reversal operator for spin one half. Hereupon 
the notation  $\phi^\pm_{}(k^\mu) = \phi^\pm$ shall be used for the sake of simplicity. 
The expression
\begin{equation}
		\s\cdot\hat{\p} \,
		\big[ \Theta (\phi^\pm)^\ast \big] = \mp \big[ \Theta (\phi^\pm)^\ast \big] 		\nonumber
\end{equation}
evinces  the helicity of $ \Theta [\phi_{}(k^\mu)]^\ast$ to be opposite to that of 
$\phi_{}(k^\mu)$, and therefore \begin{eqnarray}
		\lambda^S_\pm(p^\mu) = \sqrt{\frac{E+m}{2 m} }\left( 1\mp\frac{p}{E+m}\right)\lambda^S_\pm 		\label{jj}\\		\lambda^A_\pm(p^\mu) = \sqrt{\frac{E+m}{2 m} }\left( 1\pm\frac{p}{E+m}\right)\lambda^A_\pm
		\label{jj1}
\end{eqnarray}
 are the expansion coefficients of  a mass dimension one quantum field. In fact, the  Dirac operator $(\gamma_\mu p^\mu \pm m \mathbb{I}_4)$ does not annihilate the $\lambda(p^\mu)$ and the following results hold  \cite{allu,allu1,crevice}:
 \begin{eqnarray}
  \gamma_\mu p^\mu  \lambda^S_+(p^\mu) &=& i m  \lambda^S_-(p^\mu)    \label{1}\\ \gamma_\mu p^\mu \lambda^S_-(p^\mu) &=& - i m \lambda^S_+(p^\mu) \label{2}\\
 \gamma_\mu p^\mu \lambda^A_-(p^\mu) &=& i m \lambda^A_+(p^\mu)  \label{3}\\
 \gamma_\mu p^\mu \lambda^A_+(p^\mu) &=& - i m \lambda^A_-(p^\mu) . \label{4}
\end{eqnarray}
Nevertheless, it still implies annihilation of Elko by the Klein-Gordon operator.

Hawking radiation from black holes comprises different types of
charged and uncharged particles. We investigate
tunnelling of Elko particles from the event horizon of a black string
solution via tunnelling formalism.  By taking  
$$
\nabla_{\mu }=\partial _{\mu }+\omega _{\mu }, \,\,\,\;\;\; \omega _{\mu
}=\frac{1}{2}{}i{} \Gamma _{\mu }^{\alpha \beta }\sigma _{\alpha
\beta },$$ where $\sigma _{\alpha \beta }=\frac{1}{4}{}i{} \left[ \gamma
^{\alpha },\gamma ^{\beta }\right]$ is the spin density tensor and the $\gamma ^{\mu }$ are the usual gamma matrices satisfying the Clifford relation for Minkowski spacetime, the matrices  
\begin{equation}
\hspace*{-.2cm}\gamma ^{t}=\frac{1}{\sqrt{f}}\,\gamma ^{0}, \,\,
\gamma
^{r}=\sqrt{f}\,\gamma ^{3}, \,\,
\gamma ^{\theta }=\frac{1}{r}\gamma ^{1}, \,\,\, \gamma
^{z}=\frac{1}{\alpha r}\gamma ^{2}, \label{matt}
\end{equation}\noindent are chosen as usually \cite{ah2}, where $f=f(r)$. 
In order to find the solution of Eqs. (\ref{1})-(\ref{4}) in the background
of the black string, we employ the standard form for the Elko particle,  through the similar notation $\phi_{}^+ = \binom{\alpha}{\beta}$, where $\alpha$ and $\beta$ defined in Eq. (\ref{phim}): 
\begin{eqnarray}\label{elkof1}
\lambda^\mathrm{S}_+ \left(t,r,\theta ,z\right) &=&\left(
\begin{array}{c}
- i\beta^*  \\
i\alpha^* \\
\alpha \\
\beta
\end{array}%
\right) \exp \left( \frac{{}i{} }{\hbar }\tilde{I}\right)\,,\\
\label{elkof2}\lambda^\mathrm{S}_- \left(t,r,\theta ,z\right) &=&\left(
\begin{array}{c}
- i\alpha  \\
-i\beta \\
-\beta^* \\
\alpha^*
\end{array}%
\right) \exp \left( \frac{{}i{} }{\hbar }\tilde{I}\right)\,,\\
\label{elkof3}\lambda^\mathrm{A}_+ \left(t,r,\theta ,z\right) &=&\left(
\begin{array}{c}
i\alpha  \\
i\beta \\
-\beta^* \\
\alpha^*
\end{array}%
\right) \exp \left( \frac{{}i{} }{\hbar }\tilde{I}\right)\,,\\
\label{elkof4}\lambda^\mathrm{A}_- \left(t,r,\theta ,z\right) &=&\left(
\begin{array}{c}
- i\beta^*  \\
i\alpha^* \\
-\alpha \\
-\beta
\end{array}%
\right) \exp \left( \frac{{}i{} }{\hbar }\tilde{I}\right)\,.
\end{eqnarray}
Here   $\tilde{I}=\tilde{I}(t,r,\theta,z)$ represents the classical action. 
We use the above forms for the Elko particles   in each one of the Eqs. (\ref{1})-(\ref{4}), and solve this coupled system of  equations. Thus, by applying the WKB
approximation, where $\frac{{}i{} }{\hbar }\tilde{I} = \frac{{}i{} }{\hbar }I + I_0 + \mathcal{O}(\hbar)$, and considering terms solely up to the leading order in $\hbar $, {}{by denoting $I_{r} = {\partial I}/\partial r$, $I_{t} = {\partial I}/\partial t$, $I_{\theta} = {\partial I}/\partial \theta$, and $I_{z} = {\partial I}/\partial z$, 
this procedure yields}:
\beq
 \frac{i\alpha^* I_{t}}{\sqrt{f}}+ \beta\sqrt{f} \,I_{r} =m\beta^*+\left(\frac{i}{\alpha z} I_z - \frac{1}{r} I_\theta\right)\alpha^*\,,\label{muita} \\
 \frac{i\beta I_{t}}{\sqrt{f}}- \alpha^* \sqrt{f}\, I_{r} =m\alpha^*-\left(\frac{i}{\alpha z} I_z + \frac{1}{r} I_\theta\right)\beta^*\,.\label{muitas}\eeq\noindent
We can employ the usual ansatz in refs. \cite{k0,k1,ah2,ah4}:
\begin{equation}
I(t,r,\theta,z) =-Et+W(r)+l\theta+Jz\,, \label{nonumber}
\end{equation}
where $E$ is the energy of the emitted particles and $W$ is the part
of the action $\tilde{I}$ that contributes to the tunnelling
probability. Using this ansatz in Eqs. (\ref{muita}, \ref{muitas}) \cite{k0,k1,ah2,ah4}, the terms in (\ref{muita}, \ref{muitas}) encompassing $J$ and $l$ are dismissed.  The same solution for $J$ is obtained for both the outgoing and incoming cases. 

As it is comprehensively exposed in \cite{k0,k1,ah2,ah4},   near the black string 
horizon massive particles behave like massless particles. Phenomenologically, considering the well established Elko production by Higgs interactions  \cite{plb,plb1,plb2}, we proceed as refs.  \cite{k0,k1,ah2,ah4} and  consider the parameter $m\approxeq0$, without loss of generality, as near the horizon massive particles behave as massless ones. Thus, the
function $W(r)$ can be computed merely from Eqs.(\ref{muitas}) and (\ref{nonumber}) as:
\begin{eqnarray}
 -i \alpha^* E+\beta f(r) W'(r) =0, \nonumber\\-i \beta E+\alpha^*f(r)W'(r) =0.
\end{eqnarray}
In this case for \begin{equation}\label{++}
\alpha= i\beta^*\end{equation} we have 
\begin{equation}
W'_+(r) = E/f(r)\,,
\label{doubleumais1}
\end{equation}
\noindent whilst for the choice  \begin{equation}\label{--}
\alpha= -i\beta^*\end{equation} we get the opposite sign 
\begin{equation}
W'_-(r) = -E/f(r)\,.
\label{doubleumais2}
\end{equation}
$W_+$ [$W_-$] corresponds to outward [inward] solutions (see refs. \cite{k0,k1,ah2,ah4}). 
Eqs. (\ref{doubleumais1}) and (\ref{doubleumais2}) imply that 
\begin{equation}
W_\pm(r) =\pm \int (E/f(r))dr\,,
\end{equation}
which has a simple pole at $r=\tilde{r}$. By integrating
around the pole, it yields
\beq\label{001}
W_{\pm}(r) =\frac{\pm\pi iE}{2\alpha
^{2}\tilde{r}+\frac{M}{\tilde{r}^2}}. \eeq\noindent
The probabilities of crossing the horizon in each direction can be
given by \cite{3}
\beq
P_\pm \varpropto  e^{ -\frac{2}{\hbar}{\rm Im}\,
W_\pm(r)},\label{propto}
\eeq\noindent where $P_+$ $[P_-]$ denotes the probability of emission [absorption] by the horizon.  
While computing the imaginary part of the action, we note that it is
 the same for both the incoming and outgoing solutions. Eqs.(\ref{propto}) show now that the
probability of particles tunnelling from the inside to the outside of the
event horizon is specified by
\beq
\Gamma \varpropto \frac{P_+ }{P_-}&=&e^{ -\frac{4}{\hbar}{\rm Im}  W_{+}(r)},\eeq\noindent where in the last equality we employed Eq.(\ref{001}),  implying that  
\begin{equation}
\Gamma =\exp \left(-\frac{4\pi E}{2\alpha ^{2}\tilde{r}+\frac{M}{\tilde{r}^{2}}}%
\right) . \label{3.46}
\end{equation}
As the tunnelling probability is given by $\Gamma =\exp
\left( -\beta E\right)$, where $\beta =T_{H}^{-1}$, it yields the Hawking temperature formula
\begin{equation}
T_{H}=\frac{1}{4\pi }\left( 2\alpha
^{2}\tilde{r}+\frac{M}{\tilde{r}^{2}}\right) , \label{hawk}
\end{equation}
which is the usual Hawking temperature for black strings \cite{fatima}. Massive particles behave like massless ones and since the extra contributions vanish at the horizon, the result of integrating around the pole for $W_\pm$ in the massive case is the same as the massless case and the Hawking  temperature is recovered. Moreover, as in the Dirac tunnelling, for both the massive and massless  the Hawking temperature is obtained, {}{implying that the Elko particles  $\lambda^\mathrm{S}_+, \lambda^
 \mathrm{A}_-, \lambda^\mathrm{S}_-, \lambda^
 \mathrm{A}_+$ defined in Eqs.(\ref{pppm})-(\ref{jj1}) --- with explicit components in (\ref{elkof1})-(\ref{elkof4}) ---  are emitted at the same rate. It endows Elko particles with a different signature with respect 
to the Dirac particles (see, e. g., refs. \cite{k0,k1,ah2,ah4}), {}{which we shall emphasize below}.}

In the tunnelling formalism the probability of particles crossing the
black hole horizon on both sides are calculated using complex path
integrals.  
Solving Elko coupled equations (\ref{1})-(\ref{4}) in the background of 
black strings and by applying the WKB approximation, we have provided the 
tunnelling probability of Elko particles and the Hawking temperature associated to it.

Moreover, the tunnelling of Elko 
particles has a different feature when compared to Dirac particles. The method developed in \cite{k0,k1} for Dirac particles was further  used 
 in \cite{ah2} in the context of  black strings for the very special case where the spinor field is given by \begin{equation}
\Psi_\uparrow(t, r, \theta, z) =\left(
\begin{array}{c}
A(t, r, \theta, z) \\
0 \\
B(t, r, \theta, z) \\
0
\end{array}%
\right) \exp \left( \frac{{}i{} }{\hbar }\tilde{I}\right)\,,
\end{equation} where the author shows that there is  a 
constraint between $A$ and $B$, similarly to 
(\ref{++}) and (\ref{--}). The inward and outward 
probability of tunnelling depends on the relation 
between $A$ and $B$. For each constraint,  Dirac particles present just one behavior: either ingoing or outgoing 
particles. Notwithstanding, {}{Elko particles are 
eigenspinors of the charge conjugation 
operator, and all the eigenspinor fields ($\lambda^\mathrm{S}_+, \lambda^
 \mathrm{A}_-, \lambda^\mathrm{S}_-, \lambda^
 \mathrm{A}_+$) 
 present the same probability 
 either outgoing or ingoing for tunnelling. 
 Notice that Elko spinor field $\lambda_+^S$ in (\ref{elkof1}) differs from $\lambda_-^A$ in (\ref{elkof4}) just by the sign in the left-handed component, whereas the Elko spinor field $\lambda_-^S$ in (\ref{elkof2}) is different of $\lambda_+^A$ in (\ref{elkof3}) by the sign in the right-handed component, although they are quite different quantum fields \cite{crevice}.
 Moreover, all the four Elko particles present 
 the same inward and outward probability of 
 tunnelling and 
 the standard  Hawking temperature for black 
 strings is obtained. }

\acknowledgments
RdR is grateful to CNPq grants 303027/2012-6 and 473326/2013-2, and is also \emph{Bolsista da CAPES Proc. n$^{o}$} 10942/13-0.
 JMHS thanks to CNPq (482043/2011-3; 308623/2012-6).


\begin{thebibliography}{999}
\bibitem{1} P. Kraus and F. Wilczek,
Nucl. Phys. B \textbf{433} (1995) 403.
\bibitem{111} P. Kraus and F. Wilczek, 
Nucl.
Phys. B \textbf{437 } (1995) 231.

\bibitem{2} M. K. Parikh and F. Wilczek, 
Phys. Rev. Lett. \textbf{85} 
(2000) 5042.
\bibitem{3} K. Srinivasan and T. Padmanabhan, 
Phys. Rev. D \textbf{60} (1999) 24007.

\bibitem{4} A. J. M. Medved, 
Phys. Rev. D \textbf{66} (2002) 124009

\bibitem{5} M. Agheben, M. Nadalini, L Vanzo, and S. Zerbini,
\emph{JHEP} {\bf 05} (2005) 014.


\bibitem{6} M. Arzano, A. Medved and E. Vagenas, \ 
\emph{ JHEP} {\bf 09} (2005) 037.

\bibitem{7} Q.-Q. Jiang, S.-Q. Wu, and X. Cai,
Phys. Rev. D {\bf 73} (2006) 064003.

\bibitem{8} J. Zhang and Z. Zhao, 
Phys. Lett. B {\bf 638} (2006) 110.




\bibitem{9} P. Mitra, 
Phys. Lett. B \textbf{648} (2007) 240.

\bibitem{kernergodel} R. Kerner and R. B. Mann, 
Phys. Rev. D \textbf{75} (2007) 084022.


\bibitem{10} R. Li and J.-R. Ren, 
Phys. Lett. B {\bf 661} (2008) 370.


\bibitem{11} R. Di Criscienzo and L. Vanzo,
Europhys. Lett. {\bf 82} (2008) 60001.

\bibitem{12} R. Li and J.-R. Ren, 
Class. Quant. Grav. {\bf 25} (2008) 125016. 


\bibitem{k0}
  R.~Kerner and R.~B.~Mann,
  Phys.\ Lett.\ B {\bf 665} (2008) 277.

\bibitem{k1}
  R.~Kerner and R.~B.~Mann,
  Class.\ Quant.\ Grav.\  {\bf 25} (2008) 095014.


\bibitem{ah2}
  J.~Ahmed and K.~Saifullah,
  \emph{JCAP} {\bf 1108} (2011) 011.


\bibitem{ah4}
  U.~A.~Gillani and K.~Saifullah,
  Phys.\ Lett.\ B {\bf 699} (2011) 15.



\bibitem{vaidya} J. Ren, J. Zhang, Z. Zhao, 
Chin. Phys. Lett. {\bf 23} (2006) 2019.


 \bibitem{Banerjee:2008fz}
  {R. Banerjee  and B. R. Majhi}, 
  {}{Phys.\ Lett.\ B} {\bf 674} {(2009)} {218}.

 \bibitem{Banerjee:2008cf} 
  {}{R. Banerjee  and B. R. Majhi}, 
  {}\emph{JHEP} {\bf 06} {(2008)} {095}.

 
  
  \bibitem{Banerjee:2009wb}
  {}{R. Banerjee  and B. R. Majhi},  
  {}{Phys.\ Lett.\ B} {\bf 675} {(2009)} {243}.

\bibitem{Banerjee:2008ry}
  {}{R. Banerjee  and B. R. Majhi},  
   {}{Phys.\ Lett.\ B} {\bf 662} {(2008)} {62}.

   \bibitem{Banerjee:2010be}
  {}{R. Banerjee, B. R. Majhi and E. C. Vagenas}, 
  {}{Europhys.\ Lett.} {\bf 92} {(2010)} {20001}.

 \bibitem{Banerjee:2008sn}
  {}{R. Banerjee  and B. R. Majhi}, 
  {}{Phys.\ Rev.\ D} {\bf 79} {(2009)} {064024}.


  \bibitem{Majhi:2009xh}
  {}{B. R. Majhi}, 
  {}{Phys.\ Lett.\ B} {\bf 686} {(2010)} {49}.


  
  \bibitem{Banerjee:2009pf}
  {}{R. Banerjee, B. R. Majhi  and E. C. Vagenas}, 
  {}{Phys.\ Lett.\ B} {\bf 686} {(2010)} {279}.
  
  
\bibitem{Banerjee:2008gc}
  {}{R. Banerjee, B. R. Majhi  and S. Samanta},
  {}{Phys.\ Rev.\ D} {\bf 77} {(2008)} {124035}.
  
 \bibitem{Majhi:2009uk}
   {}{B. R. Majhi  and S. Samanta}, 
  {}{Annals Phys.} {\bf 325} {(2010)} {2410}.
  

  
  \bibitem{Majhi:2008gi}
 {}{B. R. Majhi}, 
  {}{Phys.\ Rev.\ D} {\bf 79} {(2009)} {044005}.


\bibitem{allu} D.~V.~Ahluwalia and D.~Grumiller,
  \emph{JCAP} {\bf 0507} (2005) 012.
  
  
\bibitem{horv1} {D. V. Ahluwalia \and S. P. Horvath},  \emph{JHEP} {\bf 11} {(2010)} {078}.

  
  \bibitem{allu1} D.~V.~Ahluwalia and D.~Grumiller,
  Phys.\ Rev.\ D {\bf 72} (2005) 067701.

\bibitem{al2}
  D.~V.~Ahluwalia, C.~Y.~Lee, D.~Schritt and T.~F.~Watson,
  Phys.\ Lett.\ B {\bf 687} (2010) 248.

\bibitem{al3}
  D.~V.~Ahluwalia, C.~Y.~Lee and D.~Schritt,
  Phys.\ Rev.\ D {\bf 83} (2011) 065017.

\bibitem{alex} {A. E. Bernardini and R. da  Rocha}, {Phys.\ Lett.\ B} {\bf 717} {(2012)} {238}.

\bibitem{daSilva:2012wp}
  J.~M.~Hoff da Silva and R.~da Rocha,
  Phys.\ Lett.\ B {\bf 718} (2013) 1519.


\bibitem{plb} M.~Dias, F.~de Campos and J.~M.~Hoff da Silva,
  Phys.\ Lett.\ B {\bf 706} (2012) 352.



\bibitem{plb1}  A.~Alves, F.~de Campos, M.~Dias and J.~M.~Hoff~da Silva,
  \emph{Searching for Elko dark matter spinors at the CERN LHC},
  [{\tt arXiv:1401.1127 [hep-ph]}].

\bibitem{plb2}
  B.~Agarwal, A.~C.~Nayak and R.~K.~Verma,
  \emph{ELKO as dark matter candidate},
  [{\tt arXiv:1407.0797 [hep-ph]}].  
  
 \bibitem{fab}
 {L. Fabbri}
  {Phys.\ Lett.\ B} {\bf 704} {(2011)} {255}.

 
  
  
  \bibitem{jhep} {R. da Rocha, A. E. Bernardini and J. M. Hoff da Silva}, 
  \emph{JHEP} {\bf 04} {(2011)} {110}.

\bibitem{saulo}
  {J. M. Hoff da Silva and S. H. Pereira}, 
\emph{JCAP} {\bf 03} {(2014)} {009}.

\bibitem{sh} 
 {}{A. Basak, J. R. Bhatt, S. Shankaranarayanan and K. V. Prasantha},    \emph{JCAP} {\bf 04} {(2013)} {025}.
 
\bibitem{bur1}
 {S. Kouwn, J. Lee, T. H. Lee and P. Oh}, 
  {Mod.\ Phys.\ Lett.\ A} {\bf 28} {(2013)} {1350121}.

\bibitem{bur2}
  {C. G. Boehmer, J. Burnett, D. F. Mota  and D. J. Shaw},   
  \emph{JHEP} {\bf 07} {(2010)} {053}.

  \bibitem{wund}
 {K. E. Wunderle and R. Dick}
  {Can.\ J.\ Phys.} {\bf 87} {(2009)} {909}.


  \bibitem{ro11} 
 {R. da Rocha and J. M. Hoff da Silva}, 
  {Int.\ J.\ Geom.\ Meth.\ Mod.\ Phys.} {\bf 6} {(2009)} {461}.
  
  \bibitem{ro12} 
  {}{R. da Rocha and J. G. Pereira}, 
  {Int.\ J.\ Mod.\ Phys.\ D} {\bf 16} {(2007)} {1653}.



\bibitem{horo} J. H. Horne and G. T. Horowitz, {Nucl. Phys. B} \textbf{368} (1992) 444.

\bibitem{ander} W. G. Anderson and N. Kaloper, {Phys. Rev. D} \textbf{52} (1995) 4440.

\bibitem{4658} N. Kaloper, {Phys. Rev. D}  \textbf{ 48} (1993) 4658.

 \bibitem{Chamblin:1999by}
  A.~Chamblin, S.~W.~Hawking and H.~S.~Reall,
  Phys.\ Rev.\ D {\bf 61} (2000) 065007.

\bibitem{clark} S.~S.~Seahra, C.~Clarkson, R.~Maartens,
  Phys.\ Rev.\ Lett.\  {\bf 94} (2005) 121302

 	
\bibitem{plbb}
  {}{D. Bazeia, J. M. Hoff  da Silva, R. da Rocha}, 
  {Phys.\ Lett.\ B} {\bf 721} {(2013)} {306}.
	
	\bibitem{pl1} 
  {}{R. da Rocha, A. Piloyan, A. M. Kuerten,  C.~H.~Coimbra-Araujo}, 
  {Class.\ Quant.\ Grav.} {\bf 30} ({2013})  {045014}.


 \bibitem{pl2} 
 {}{R. da Rocha, J. M. Hoff da Silva}
  {Eur.\ Phys.\ J.\ C} {\bf 72} {(2012)} {2258}.











\bibitem{lemos95a} J. P. S. Lemos, {Class. Quant. Grav.} \textbf{12} (1995) 1081.

\bibitem{LZ} J. P. S. Lemos and V. T. Zanchin, {Phys. Rev. D} \textbf{54} (1996) 3840.


\bibitem{S} N. O. Santos, {Class. Quant. Grav.} \textbf{10} (1993) 2401.


\bibitem{crevice} D.~V.~Ahluwalia,
\emph{On a local mass dimension one Fermi field of spin one-half and the theoretical crevice that allows it} [{\tt arXiv:1305.7509 [hep-th]}].




\bibitem{fatima} A. Fatima and K. Saifullah, 
Astrophys. Space Sci. {\bf 341}  (2012) 437.
\bibitem{CZ} R. G. Cai and Y. Zhang, {Phys. Rev.} D \textbf{54} (1996) 4891.

\end{thebibliography}
\end{document}